# Multi-level Electro-thermal Switching of Optical Phase-Change Materials Using Graphene

Carlos Ríos[1,*], Yifei Zhang[1], Mikhail Shalaginov[1], Skylar Deckoff-Jones[1], Haozhe Wang[1], Sensong An[2], Hualiang Zhang[2], Myungkoo Kang[3], Kathleen A. Richardson[3], Christopher Roberts[4], Jeffrey Chou[4], Vladimir Liberman[4], Steven A. Vitale[4], Jing Kong[1], Tian Gu[1], Juejun Hu[1,*]

[1] Department of Materials Science & Engineering, Massachusetts Institute of Technology, Cambridge, MA, USA
[2] Department of Electrical & Computer Engineering, University of Massachusetts Lowell, Lowell, MA, USA
[3] The College of Optics & Photonics, Department of Materials Science and Engineering, University of Central Florida, Orlando, FL, USA
[4] Lincoln Laboratory, Massachusetts Institute of Technology, Lexington, MA, USA

* carios@mit.edu, hujuejun@mit.edu

**Abstract**

Reconfigurable photonic systems featuring minimal power consumption are crucial for integrated optical devices in real-world technology. Current active devices available in foundries, however, use volatile methods to modulate light, requiring a constant supply of power and significant form factors. Essential aspects to overcoming these issues are the development of nonvolatile optical reconfiguration techniques which are compatible with on-chip integration with different photonic platforms and do not disrupt their optical performances. In this paper, a solution is demonstrated using an optoelectronic framework for nonvolatile tunable photonics that employs undoped-graphene microheaters to thermally and reversibly switch the optical phase-change material $Ge_2Sb_2Se_4Te_1$ (GSST). An *in-situ* Raman spectroscopy method is utilized to demonstrate, in real-time, reversible switching between four different levels of crystallinity. Moreover, a 3D computational model is developed to precisely interpret the switching characteristics, and to quantify the impact of current saturation on power dissipation, thermal diffusion, and switching speed. This model is used to inform the design of nonvolatile active photonic devices; namely, broadband $Si_3N_4$ integrated photonic circuits with small form-factor modulators and reconfigurable metasurfaces displaying $2\pi$ phase coverage through neural-network-designed GSST meta-atoms. This framework will enable scalable, low-loss nonvolatile applications across a diverse range of photonics platforms.

**Keywords:** graphene microdevices, optical phase-change materials, nonvolatile photonics, optoelectronics

## 1. Introduction

The ability to reconfigure optical devices to adapt to different tasks on-the-fly has been a long-sought-after goal with profound impacts on applications including optical communications,[1,2] quantum optics,[3,4] analog computing,[5–8] active metasurfaces,[9–11] device trimming,[12–14] and photonic sensing.[15,16] These devices are often implemented using thermo-optic, electro-optic, or all-optical actuation mechanisms, which demand constant power supply to maintain the optical state. The volatility is far from ideal for applications where the optical configurations only need to be changed sporadically. For these applications, nonvolatile reconfiguration capable of retaining any configuration with zero-power consumption is the key functionality.[17] Thus far, nonvolatile reconfiguration has been achieved in integrated optics using charge trapping effects,[18,19] nanomechanical resonators,[20] ferroelectric materials,[21] and phase-change materials (PCMs).[17] PCMs allow for complex refractive index modulation by using thermal stimuli to switch between the fully amorphous and the fully crystalline states, and also between any intermediate mixture level.[22] The giant optical property contrast between the amorphous and crystalline states of PCMs has been exploited to demonstrate low-energy integrated optical switches,[23–27] multilevel memories,[22,28] reconfigurable metasurfaces,[29–32] color pixels,[33] and building blocks for brain-inspired computing.[5,34–36] Besides novel applications, there has also been a drive towards finding new phase-change materials designed and optimized for optical applications,[37,38] epitomized by the broadband transparent PCM $Ge_2Sb_2Se_4Te_1$ (GSST).[39] Another important aspect is the development of switching mechanisms that enable reconfiguration of PCMs without adversely impacting the device performance. In particular, electro-thermal methods that enable scalable on-chip integration have been explored in several recent studies.[39–44] The heater materials used include metals,[45] transparent conducting oxides (TCOs),[46–49] and doped silicon. While metals are useful for free-space reflective devices, they introduce significant optical losses in transmissive or waveguide components. Doped silicon is an ideal choice for PCM integration with the silicon-on-insulator platform. However, it is challenging to apply it to $Si_3N_4$-based devices—another widely deployed photonic platform, or other non-silicon waveguide platforms. Moreover, when it comes to metasurfaces, the design PCM or hybrid meta-atoms with broad phase modulation is non-trivial when considering a silicon slab or substrate, given that silicon's large refractive index favors confined modes in the heater itself. TCO heaters, while useful for devices operating in the visible spectrum, suffer from elevated optical losses in the infrared due to free carrier absorption.

To simultaneously resolve these issues, graphene is put forward as the best option given its high thermal and electrical conductivity, integration versatility, and superior stability.[50,51] In addition, infrared optical losses of graphene can be minimized by harnessing the doping-induced Pauli blocking effect.[52] According to a recent theoretical analysis,[53] graphene heaters boast two orders of magnitude higher figure of merits for heating and overall performance compared to Si or TCO heaters when applied to PCM switching. Electro-thermal devices using graphene have been demonstrated in optical modulation applications based on black-body broadband radiation.[54,55] Using a similar device with single-layer graphene, we can use the high temperatures to switch a PCM cell placed directly on top of the heater.

Here we have demonstrated, for the first time, reversible switching of PCM by using single-layer graphene microheaters. We further realize switching between four different crystallinity levels using these devices. Moreover, we build a 3D computational model to reproduce the experimental outcomes, and thus, allow analysis of the influence of the thermal boundary conductivity between graphene and $SiO_2$. In the final two sections, we leverage the calibrated computational model to design $Si_3N_4$ (SiN) integrated photonic devices with ultra-compact phase shifters, and reconfigurable metasurfaces with full $2\pi$ optical phase coverage.

2. **Experimental Results**

Our devices consist of single-layer graphene transferred onto 3-μm-thick $SiO_2$/Si samples and patterned into the microheater shape following the fabrication process described in the *Experimental Section*. We chose GSST as the PCM, which was deposited using thermal evaporation and patterned via a lift-off process. Two Ti/Au metal pads were used as electrical contacts, and $Al_2O_3$ films grown by atomic layer deposition served as conformal protective layers. A cross-section sketch of the device is shown in **Figure 1a,** together with the equivalent electrical circuit of the device. The chip was wire bonded to a custom printed circuit board (PCB) and mounted on a 3D printed stage to make it compatible with *in-situ* Raman measurement – a photograph of the device is shown in Figure 1b. Using Raman spectroscopy characterization, we measured the primary in-plane vibrational mode, denoted as G peak, at ~1587 cm$^{-1}$ and the maximum of the 2D peak of graphene across all devices at ~2704 cm$^{-1}$. We attribute the shift of this peak from the typical 2690 $cm^{-1}$ to p-doping from the $SiO_2$ substrate – especially after alumina deposition at high temperatures [56,57] – and to compressive strain from the several fabrication steps. With an intensity ratio $I_{2D}/I_G$ of up to 4.21, we verified that our graphene consisted of a single-layer sheet. An optical microscope image of a device top view is shown in Figure 1c, which shows the microheater consisting of the two Ti/Au metallic pads in contact with a 100 μm wide graphene, featuring a 5×w μm² bridge in the center of the device. We fabricated devices with the bridge width $w = 3,5,10$ μm, and GSST cells of various areas: 3 × 4 μm² and 1.5 × 4 μm² for $w = 5,10$ μm, and $w = 3$ μm, respectively. Figure 1d shows the characteristic *I-V* curves for the three type of devices tested in this work, which display a quasi-ohmic regime followed by a current saturation plateau. This phenomenon is well known for undoped graphene microheaters with a few to tens of microns in size, and situated on a SiO₂, a polar substrate. When voltage bias is applied between the two electrodes, a hotspot is created in the Dirac point of the lattice and high temperatures can be reached due to Joule heating. Additionally, surface polar phonons (SPoPh) are generated at the graphene/SiO₂ interface, which increase the electron scattering and reduce the overall charge mobility.[58–60] This scattering increases with current (voltage) and becomes relevant in the devices with dimensions comparable to the SPoPH propagation length, which is ~10 μm.[61,62] Given the geometries of our devices, the electron scattering effect is significant at high voltages, creating a bottleneck for the electron current in the graphene bridge. Consequently, the SPoPh scattering also affects the thermal conductivity between graphene and the SiO₂ substrate by dissipating heat in-plane, and thus, decreasing the thermal conductivity towards SiO₂, which will be detailed in the following section. This effect, in turn, means that high temperatures in the GSST cell are reached with exceptionally low power consumption, but with more prolonged heating and cooling time constants, commensurate with the micro- to millisecond switching times of GSST.[39,63] We note that the *I-V* curves were very close among the same type of microheaters, thus showing that the current saturation is an intrinsic property of the geometry, as opposed to a random effect. We display only one full 0-10V *I-V* for $w = 3$ μm, since the device was broken afterwards; other identical devices displayed a damage threshold of ~7 V, which is in good agreement with similar devices.[64] On the contrary, devices with $w= 5$ μm and $w = 10$ μm performed well even after driving with up to 10 V.

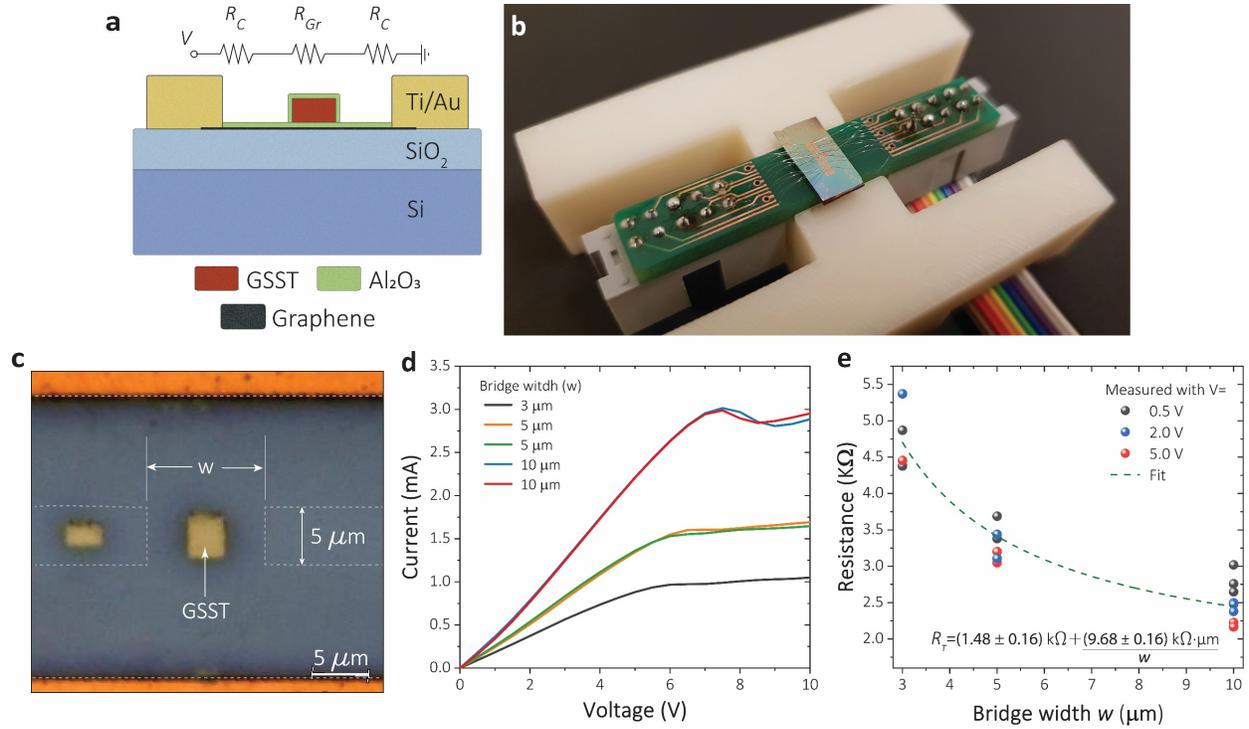

**Figure 1. Graphene microheater characterization. a** Sketch of the transversal section of the device comprising a graphene microheater and a GSST cell. Thin layers of alumina are used to protect both graphene and GSST from the environment. The equivalent circuit is also plotted, where $R_C$ is the contact resistance, $R_{Gr} = 2 \cdot R_{Sh}(10.5\ \mu m/100\ \mu m) + R_{Sh}(5\ \mu m/w)$ is the total graphene resistance as a function of the sheet resistance, $R_{Sh}$, and the bridge width $w$. **b** Photograph of the chip wire bonded to a custom printed circuit board (PCB) and mounted on a 3D printed stage **c** Top-view optical microscope image of a device with $w = 10$ μm, and 50 nm thick GSST placed on the microheater bridge. The dotted lines delimit the graphene. The GSST cell outside the graphene microheater was used as reference. **d** Current-voltage (*I-V*) curves of five different devices with $w$=3,5 and 10 μm, displaying a nearly ohmic behavior and then a saturation effect. **e** Total resistance of the devices in **c** extracted at 0.5, 2 and 5 V, in the quasi-linear regime. The fitting equation for the concatenated data is shown. From this fitting equation we calculate the approximated experimental values of $R_C$ and $R_{Sh}$.

We extract the sheet ($R_{Sh}$) and contact ($R_C$) resistance of our devices by plotting the total resistance $R_T = 2 \cdot R_C + 2 \cdot R_{Sh}(10.5\ \mu m/100\ \mu m) + R_{Sh}(5\ \mu m/w)$ as a function of bridge width $w$ at three different voltages within the quasi-ohmic regime. The results are shown in Figure 1e. The total resistance is independent of voltage within our measurement accuracy. Assuming a negligible variation of the total resistance, we parameterized the resistance contributions by fitting the experimental data. The fitted curve is a good approximation based on the comparison between experimental and simulation results to be studied later in this paper. By comparing the fitting function displayed in Figure 1e, and the total resistance, we calculated $R_{Sh} = 1936 \pm 16\ \Omega$ and $R_C = 569 \pm 82\ \Omega$.

We demonstrate reversible and controllable switching in devices with 50-nm-thick, $3 \times 4\ \mu m^2$ GSST cells using a graphene microheater with $w = 10$ μm (see Figure 1). To crystallize the GSST cell (i.e., to heat up above the crystallization temperature $T_c \approx 550$ K), we used 6 V and 20-ms-long pulses, with a 1.5 ms trailing edge. To amorphize (i.e., to heat up over the melting temperature $T_m \approx 890$ K and quench), we used 7.5 V and 13-μs-long electrical pulses – right at the saturation point of the device, as shown in Figure 1d. The pulse sequence for reversible switching is sketched in **Figure 2a.** With a total resistance of 2.51 ± 0.05 kΩ, measured at 7.5 V, the device consumed a total of 22.4 ± 0.6 mW, of which 8.6 ± 1.1 mW was dissipated across the 10 × 5 μm² microheater bridge. We denote the power required by the graphene microheater bridge to switch to amorphize the entire GSST cell, in this case, 8.6 ± 1.1 mW, by $P_{Am}$. Given a fixed heater geometry, $P_{Am}$ is independent of the contact or sheet resistance. The total energy consumption to switch to the amorphous state is 111.8 ± 14.3 nJ. Similarly, 14.3 ± 0.3 mW was the total power to crystallize, of which 5.52 ± 0.7 mW was dissipated by the graphene bridge, which we denote as $P_{Cry}$. The total energy for a crystallization pulse was of 110.4 ± 14.0 μJ. The remaining power was lost in the graphene pads and, mostly, at the graphene/metal boundary due to considerable contact resistance. The power absorbed outside of the bridge can be suppressed by further device engineering.

A sequence of the two pulses shown in Figure 2a allowed us to switch between four different crystallization levels, which we demonstrate in Figure 2b. To do so, we sent either one or two amorphization pulses (blue) to reach two distinct predominantly amorphous states, and two or three crystallization pulses (red) to reach two different predominantly crystalline states – except in the first crystallization event, in which we also tested pulses shorter than 20 ms. Using an *in-situ* Raman testing setup (see *Experimental Section* and Ref. [65]), we were able to track the Raman signal of both the amorphous (159 cm$^{-1}$) and the crystalline (120 cm$^{-1}$) signature peaks after each switching event. By measuring the difference between the normalized Raman signal (NRS) for each peak, we observed reversible and controllable switching between the four distinguishable levels without device damage during the entire experiment. In Figure 2c, we show the Raman spectra for the four data points highlighted in Figure 2b corresponding to the data points in each of the four different levels. We attribute the variations within each level to the stochastic nature of nuclei generation induced by heat transfer from the microheater [66] and to the time and space fluctuations of the graphene hotspots through a random scattering of surface polar phonons when operating the device at its saturation point.[58] Future research is warranted to elucidate the origin of the fluctuations. Our results represent a step toward the development of broadband transparent and substrate-agnostic microheaters to electro-thermally switch phase-change photonic devices. In the following sections, we develop a comprehensive computational model to elucidate the operation mechanisms of the graphene microheaters, and further analyze the integration of these graphene microheaters with on-chip devices to enable scalable, low-loss, and low-energy phase shifters for optical routing and computing, as well as reconfigurable metasurfaces with individual meta-atom tunability.

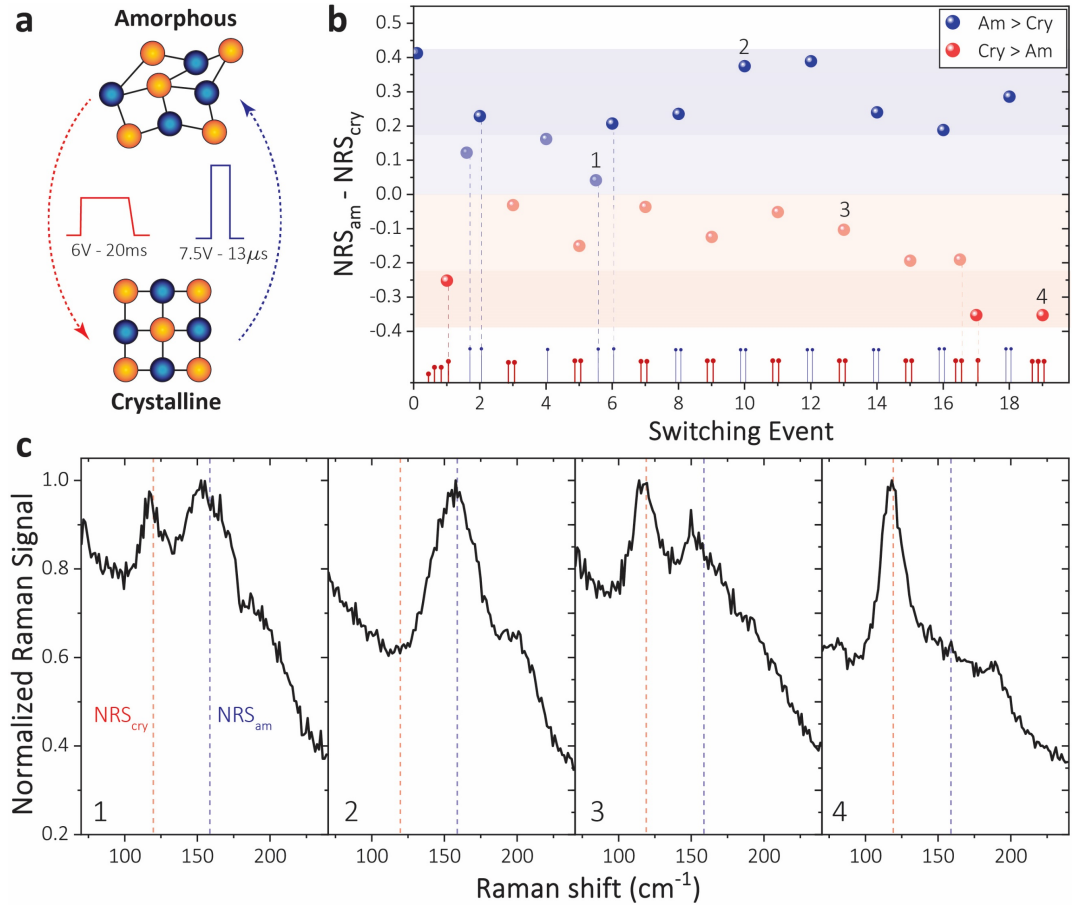

**Figure 2. *In-situ* switching and Raman spectroscopy demonstrating reversible multi-level switching. a** Sketch of the mechanism to electro-thermally switch between the amorphous and crystalline states of GSST. The values for the pulse duration and amplitude correspond to those used to switch a device with $w=10$ μm. **b** Demonstration of reversible switching by measuring the peak height difference between the normalized Raman signals (NRS) for the amorphous and the crystalline peaks at 159 cm$^{-1}$ and 120 cm$^{-1}$, respectively. The pulse sequence triggering the transitions is plotted at the bottom. Each red pulse corresponds to the crystallization pulse shown in **a,** with a total energy of 110.4 ± 14.0 μJ. Similarly, each blue pulse, for amorphization, has a total energy of 111.8 ± 14.3 nJ. Except for the first crystallization event in which four pulses were used (same power but lengths of 10 ms, 2 × 15 ms, and one of 20 ms for a total of 331.2 ± 24.2 μJ), either two (220.8 ± 19.8 μJ) or three (also with 331.2 ± 24.2 μJ) pulses were used to reach two distinct predominantly crystalline levels, highlighted in red. Either one (111.8 ± 14.3 nJ) or two (223.6 ± 20.2 nJ) amorphization pulses were used to reach the two distinct predominantly amorphous states, highlighted in blue. **c** Raman spectra for the four points highlighted in **b**. The red and blue dotted lines indicate the Raman peaks corresponding to the amorphous and the crystalline states.

## 3. Multi-physics Simulation Model

We constructed a full 3D computational model in COMSOL Multiphysics, using the modules and material properties described in the *Experimental Section.* We consider every material, dimension, and electrical pulse property employed in the experimental demonstration to reproduce as accurately as possible the results in Figure 2. The most important parameter to tune in the model is the thermal conductivity between graphene and the SiO$_2$ substrate, $h_{SiO/Gr}$. This parameter is relevant to understand the power dissipation given that $P = V^2/R \approx h_{SiO/Gr} \cdot A \cdot (T - T_0)$, where *V* denotes the applied voltage, *R* gives the electrical resistance of the heater, *A* is the area of the boundary, *T* is the heater temperature, and $T_0$ is the ambient temperature. For constant power, the higher the $h_{SiO/Gr}$ value, the lower the final *T* as more heat is conducted to the substrate. To avoid further increasing the complexity of our model, we approximate the two-dimensional electric phenomena at high, constant voltage (i.e. charge mobility, SPoPh scattering, and phonon transport) to a single effect: the modulation of the thermal boundary resistance between graphene and SiO$_2$, $h_{SiO/Gr}$. This approximation, as we will show in this paper, is adequate to reproduce with high accuracy our experimental data.

Using the pulse parameters that experimentally triggered uniform switching of the entire 50-nm-thick GSST cell, we calculate the values for $h_{SiO/Gr}$ such that the melting temperature is reached using a 7.5 V and 13 µs pulse. The results are shown in **Figure 3a**. We found that $h_{SiO/Gr} = 1.8 \times 10^5$ W m$^{-2}$ K$^{-1}$ is required to first reach an average $T = T_m$, but only $h_{SiO/Gr} = 1.5 \times 10^5$ W m$^{-2}$ K$^{-1}$ guarantees a uniform switching of the entire GSST cell, i.e., that all the GSST is above 890 K – taking as a reference Ge$_2$Sb$_2$Te$_5$ (GST) amorphization temperature.[67] The latter result we obtained here is one order of magnitude smaller than the one found in similar devices in Ref. [68] However, other works reported thermal boundary conductivities between graphene and SiO$_2$ as high ~10$^8$ W m$^{-2}$ K$^{-1}$ and as low as ~10$^3$ W m$^{-2}$ K$^{-1}$,[69] displaying up to five orders of magnitude decrease in thermal resistance as a result of graphene corrugation, for instance. In our case, we attribute the low values to the operation of the graphene microheater at the current saturation point, i.e., at maximum SPoPh generation and electron-SPoPh scattering. The electrical and thermal dynamics at the boundary between graphene and SiO$_2$ depend highly on the generation of SPoPh and the electron scattering they introduce. It is well known that the surface polar phonons at high voltages modulate the heat dissipation towards the substrate by modifying $h_{SiO/Gr}$, a phenomenon extensively studied in Ref. [68]. Since the pulse excitation for both amorphization and crystallization happens at high voltages, we expect the SPoPh effect to be predominant, especially for amorphization, which is triggered with voltages close to the saturation point. Besides electronic and phononic effects, single-layer graphene quality and its mechanical contact with the substrate can also impact $h_{Gr/SiO2}$. Several defects can lead to a variation of the thermal transport properties of graphene, namely, contaminants in the SiO$_2$ substrate, vacancies, grain boundaries, Stone–Wales defects, substitutional and functionalization defects, and wrinkles or folds.[70] Additionally, the melting temperature of GSST is not yet characterized, which is the reason why we use the GST value as a reference. Previous results suggest that a similar compound, Ge$_2$Sb$_2$Se$_{4.5}$Te$_{0.5}$, has a melting temperature of approximately 730 K,[71] which is significantly lower and might be another reason why we get a small value for $h_{SiO/Gr}$. This melting temperature, in Figure 3a, would lead to a higher $h_{SiO/Gr} = 5 \times 10^5$ W m$^{-2}$ K$^{-1}$, in which case our simulations remain qualitatively the same, except that the temperature scales would reflect a $T_m \approx 730$ K instead of $T_m \approx 890$ K.

Using $h_{SiO/Gr} = 1.5 \times 10^5$ W m$^{-2}$ K$^{-1}$, we simulated the average temperature at the bottom surface of the GSST cell as a function of time, for both crystallization and amorphization pulses. Figure 3b shows the results for the three microheater geometries operated at the maximum achievable current values (see Figure 1) with 7.5 V, 13 µs (red curves) and 6 V, 20 ms (black curve) pulses for enabling amorphization and crystallization. For the $w = 10$ µm devices, we found that both simulated pulses can heat GSST cell over the melting and crystallization temperatures. Similarly, Figure 3c shows the cooling curve for the same device, which takes less than 2 µs to cool down below the crystallization temperature, a quenching process that guarantees amorphization in GSST.[63] This cooling process was best fitted using a double exponential function, displaying two thermal constants: $\tau_1 = 4.91 \pm 0.19$ µs and $\tau_2 = 0.708 \pm 0.006$ µs, which is consistent with experimental results observed before.[72] We attribute this effect to the different rates of thermal dissipation taking place in and out of plane. Initially, the heat dissipation throughout the graphene sheet and towards the metal contact dominates and takes place at the fast time scales – simulating the electronic and phononic effects in single-layer graphene. Subsequently, when in-plane 'equilibrium' is reached, the heat dissipation towards the substrate takes over, driving a slower cooling down process. This effect demonstrates the need of a full 3D model in the study of these graphene microheater devices to accurately capture thermal dynamics and properly inform experimental tests.

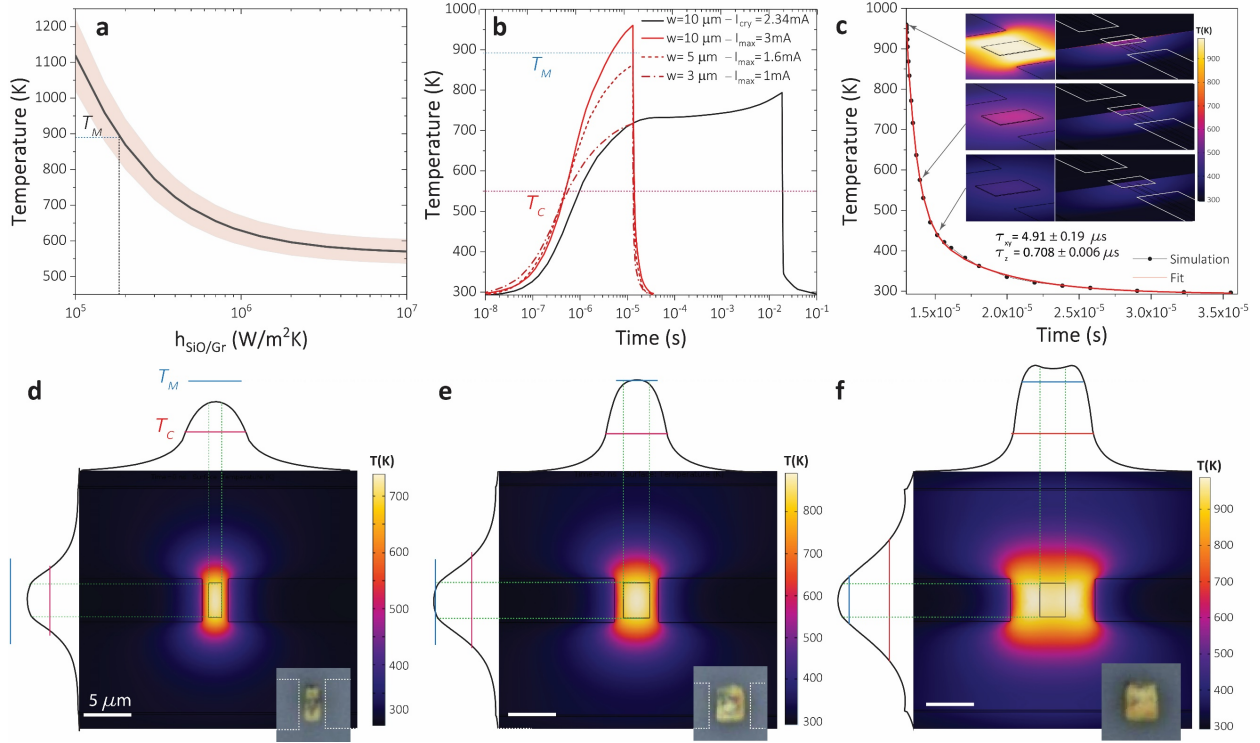

**Figure 3. 3D computational simulation of graphene microheater devices with GSST cells. a.** Maximum temperature at the bottom surface of the GSST cell, as a function of $h_{SiO/Gr}$, thermal boundary conductivity between graphene and SiO$_2$ substrate. We used a graphene microheater bridge with $w$ = 10 µm, and a pulse with 7.5 V amplitude and 13 µs duration. We found that $h_{SiO/Gr} = 1.5 \times 10^5$ $W\,m^{-2}K^{-1}$ is required to achieve full-area switching. **b.** Temperature evolution during pulse excitation for devices with $w$ = 3, 5 and 10 µm using the maximum current at the saturation point (see Figure 1c) and $h_{SiO/Gr} = 1.5 \times 10^5$ W m$^{-2}$ K$^{-1}$. The 7.5 V and 13 µs pulse used in the experiment (see Figure 2a), for $w$ = 10 µm devices, led to temperatures over GSST melting point. The 6 V and 20 ms pulse reached a maximum temperature of 790 K, enough to crystallize without reaching melting temperature, which replicates our experimental switching conditions. **c.** Cooling curve for the amorphization pulse applied to the $w$ = 10 µm device in **b**. A double exponential with thermal constants $\tau_1$ = 4.91 ± 0.19 µs and $\tau_2$ = 0.708 ± 0.006 µs provides the best fit to the simulation results. The insets show the temperature profiles on the top surface (left column) and a plane normal to the surface (right column) to illustrate in-plane (throughout graphene) and perpendicular (towards the substrate) heat dissipation, respectively. **d-f** 2D maximum temperature profile at the end of each 13 µs pulse in **b**, for the three devices under consideration. $T_c$ (red line) and $T_m$ (blue line) are the crystallization and the amorphization temperatures, respectively, plotted to illustrate the range of temperatures reached by each device. The insets show the optical microscope images for each of the devices tested experimentally. The inset in **d** displays a central oval-shaped gradient in crystallinity (noted as a difference in color), elongated in the direction of the bridge. This agrees well with the simulated temperature gradient profile. The inset in **e** shows a small central area that was experimentally switched back and forth. This is consistent with the temperature profiles where the center area of the GSST cell does reach melting temperatures, yet the periphery regions remain below $T_m$, as shown in **b**. The inset in **f** shows a device after 15 switching events. The uniformity of color throughout the GSST cell is the consequence of the entire cell reaching temperatures over $T_m$, which is accurately reproduced in the simulated device.

Given the restrictions imposed by the damage threshold, the devices with $w$ = 3 µm cannot reach melting temperatures, as shown in Figure 3b. The same conclusion can be made from a top-side-view thermal map of the device at the end of the 13 µs pulse (Figure 3d). This result agrees well with the *in-situ* Raman measurements in this type of devices, which only crystallized but did not amorphize. Figure 3d shows the temperature profile inside the GSST cell with a prominent gradient within the 3 µm-wide bridge, forming an elliptical area (from the superposition of the two gaussian-like profiles) where higher temperatures were observed. The optical microscope image (inset in Figure 3d) confirms the simulation outcomes: the entire cell was crystallized, but the center displayed an oval shape in which the level of crystallinity is higher.

For devices with $w = 5$ µm, the current saturation drives the cell on the verge of full cyclability. With a current equal to 1.6 mA, as shown in Figure 3b, the pulse does not elevate the temperature over $T_m$; however, this is an average over the bottom surface of the GSST cell. In Figure 3e, we show the 2D top view temperature profile at the end of the 13 µs pulse. In this figure, we see how only an elliptical area within the cell barely reaches amorphization temperatures. This result matches well our experimental results if compared to the optical microscope image inset, where we show the inner area that cycled back and forth. For the devices with $w = 10$ µm, we can see from Figure 3b and 3f that the entire GSST cell, which lies inside the area that can reach $T_m$, can be cycled back and forth uniformly. In the microscope image inset, one can see that after 15 switching events, the GSST surface remains almost uniform. Our 3D model, therefore, accurately reproduces the results observed in our experiment with the calibrated $h_{SiO/Gr}$ value. Hence, we proceed to use this same model to inform the design of active photonic devices based on graphene microheaters and GSST.

### 3.1 Nonvolatile Electro-thermal Modulation of $Si_3N_4$ Photonic Integrated Circuits

SiN photonic integrated circuits – a low-cost, low-loss, broadband transparent and nonlinear platform – have been widely employed for applications such as communications, beam steering, sensing, signal processing, and optical gyroscopy.[73] PCMs present a promising solution to impart nonvolatile control capabilities to the SiN platform. Here, we propose graphene microheaters to electro-thermally switch PCMs integrated onto $Si_3N_4$ waveguides with ultra-low power consumption and minimal insertion loss. Since both graphene and SiN display broadband transparency, we envision devices that can be operated at wavelengths as short as 500 nm.[74] We study the device sketched in **Figure 4a**, which is based on the device we tested experimentally (see Figure 1). This device consists of a 400-nm-thick and 800-nm-wide $Si_3N_4$ waveguide on $SiO_2$ substrate, further planarized to create a flat surface to which the graphene is transferred. The planarization prevents the graphene from rupturing or introducing undesired strain that in turn, modify the electrical conductivity properties of graphene.[75] The single-layer graphene was simulated with a chemical potential of 0.2 eV, which was experimentally measured at room temperature [62,76] and reflected the doping of graphene by the SiO2 substrate.[56] In Figure 4b, we plot the calculated real part of the effective refractive index for 30-nm-thick GSST covering the entire 800-nm-wide waveguide. We observe a change in the real effective refractive index of $\Delta n_{eff}$ = 0.28 and $\Delta n_{eff}$ = 0.064, at λ = 1550 nm, for the two fundamental modes, TE and TM, respectively. The much smaller effective index change for the TM mode is attributable to an 'anti-slot' effect which diminishes field confinement in the high-index PCM region. The index modulation is three to four orders of magnitude larger than the thermo-optical effect used in Si3N4 modulators;[77] which in practice means that devices require smaller form factors to reach, for instance, a π phase shift. Thicker GSST offers a larger modulation of the refractive index at the cost of higher losses. Besides, thicker crystalline GSST gives rise to a family of higher-order TE modes confined within the GSST rather than in the Si3N4 waveguide, given its higher refractive index. For 30-nm-thick crystalline GSST and wavelengths shorter than 1.43 µm, for instance, the devices operating with TE modes display an effective refractive index higher than SiN refractive index, as shown in Figure 4b. The strong coupling also leads to higher losses, which are shown in Figure 4c for the TE and TM modes and for GSST in both states. We observe that the TE modes undergo higher losses when propagating through the GSST device, displaying a trade-off between large $\Delta n_{eff}$ and total loss. To study this trade off, we propose the following figure of merit: $FOM = \Delta n_{eff}^2 / (P_\pi \cdot k_{eff,cry}) \propto (P_\pi \cdot L_\pi \cdot IL)^{-1}$, where $L_\pi$ denotes the device length to obtain a π phase shift, $IL$ is the total insertion loss, and $P_\pi$ represents power to switch the entire GSST cell. The results in Table 1 show that the 30 nm GSST and TE mode reach the highest FOM due to the shortest $L_\pi$ and therefore the smallest $P_\pi$. We also find that the minimum $IL$ for a π phase shift is achieved with a 20 nm thick GSST and the TM mode, given the low $k_{eff,cry}$. However, for this thickness and polarization, $L_\pi$ = 19.4 µm which requires a significantly larger graphene device and switching power. Additionally, in Figure 4c we plot the losses of the same device without GSST to illustrate the low insertion loss of graphene, which is fully transparent in the wavelength range studied here. At λ = 1550 nm, the single-layer graphene introduces losses of 0.05 dB and 0.03 dB for the TE and TM modes, respectively. Furthermore, if we consider a higher chemical potential, 0.34 eV which is in the range of values obtained with polar substrates,[62] the losses are reduced to

0.03 dB for TE and 0.02 dB for TM, making the graphene even more transparent. These results are in good agreement with the experimental measurements in Ref. [78]

Table 1. Phase modulation using graphene microheaters to control GSST with three different thicknesses

| GSST Thick (nm) | TE | | | | | | TM | | | | | |
|---|---|---|---|---|---|---|---|---|---|---|---|---|
| | $\Delta n_{eff}$ | $k_{eff\,cry}$ | $L_\pi$ (µm) | IL (dB) | $P_\pi \approx P_{Am}$ (mW) | FOM | $\Delta n_{eff}$ | $k_{eff,cry}$ | $L_\pi$ (µm) | IL (dB) | $P_\pi \approx P_{Am}$ (mW) | FOM |
| 10 | 0.044 | 0.0166 | 17.6 | 10.37 | 18.97 [2] | 6.2 | 0.021 | 0.006 | 37.0 | 7.85 | 30.98 [3] | 2.4 |
| 20 | 0.125 | 0.0512 | 6.2 | 11.17 | 8.6±1.1 [1] | 35.5 | 0.04 | 0.0101 | 19.4 | **6.92** | 18.97 [2] | 8.4 |
| 30 | 0.28 | 0.146 | **2.77** | 14.2 | 8.6±1.1 [1] | **62.4** | 0.064 | 0.017 | 12.1 | 7.29 | 13.65 [4] | 17.7 |

[1] Experimental result for the 10 × 5 µm² microheater studied so far (see Figure 4d), which can be used to switch both a 2.77 µm (using a 5 µs pulse) and a 6.2 µm (using a 13 µs pulse) long GSST cells.
[2] From the simulation of a 25 × 5 µm² microheater. [3] 45 × 5 µm² wide microheater. [4] 15 × 5 µm² µm wide microheater.

In Figure 4d, we show the top view of the temperature profile right after a 7.5 V and 5 µs pulse employing a 5 µm wide and 10 µm long graphene microheater with a 30 nm thick GSST cell of area 0.8 × 3 µm² – the device with the highest FOM in Table 1. Although the microheater bridge can be of smaller size, we choose the dimensions of the heater that ensure uniform reversible switching in Figure 2 and Figure 3, which overcomes the current saturation limitations of undoped graphene. Figure 4e shows 7.5 V pulses with two different durations, 13 µs and 5 µs. The former is the pulse width used in the experimental results in Figure 2, and the latter is the shortest pulse that can uniformly switch the GSST cell on top of the waveguide, both dissipating 8.6 mW. Since $Si_3N_4$ is a better thermal conductor than $SiO_2$, the temperature rises faster upon the pulse incidence; this effect enables using 5 µs pulses instead and reducing the amorphization energy consumption from 111.8 nJ to 43 nJ. The crystallization pulse remains the same as in Figure 3, although microsecond pulses can trigger crystallization, the long duration of 20 ms pulses permits to reach higher crystalline states. Heat dissipation in the nanosecond regime, of interest for faster crystallization kinetics PCMs, is not feasible under the current experimental conditions. The significant contrast of thermal conductivity between graphene and $SiO_2$ implies that 4-5 times larger power dissipation needs to happen on the microheater to reach $T_m$ in less than 100 ns. Nevertheless, the current saturation effect in low dimensional graphene precludes increasing the current beyond the levels experimentally tested here for single-layer graphene. Besides, the contact resistance needs to be significantly lowered to avoid heat dissipation in the electrode-graphene boundary, which can cause breakdown of the device.

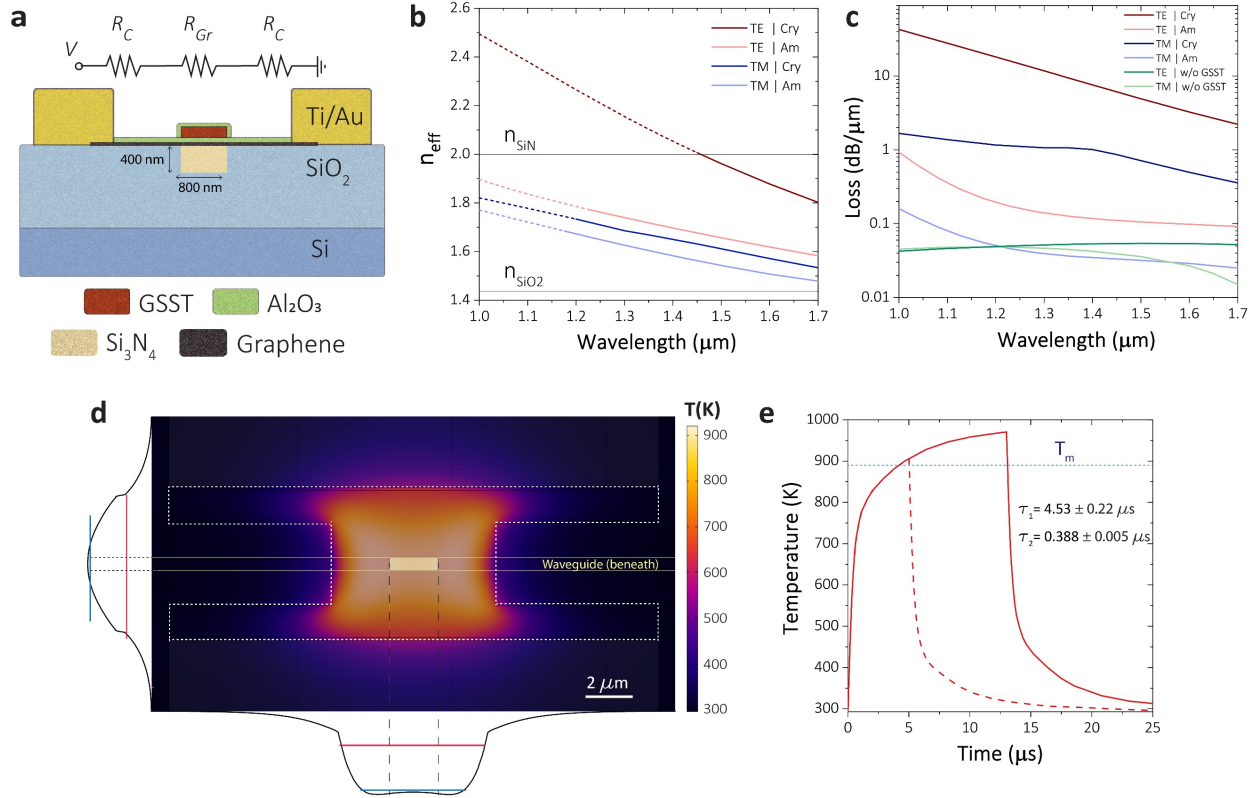

**Figure 4**. **Si$_3$N$_4$ integrated photonic circuit based on GSST. a.** Cross-section sketch of the planarized 800 × 400 nm$^2$ Si$_3$N$_4$ waveguide with a graphene microheater and a GSST cell. **b**. Real part of the effective refractive index as a function of wavelength. The TE and TM fundamental modes were calculated using the finite element method in Lumerical Mode. The dotted lines represent higher-order modes, while the solid lines represent single-mode operation. **c**. Propagation losses of the TE and TM fundamental modes, calculated from the imaginary part of the effective refractive index, $k_{eff}$, as $\alpha$(dB· µm$^{-1}$)=4.34·4π·$k_{eff}$/$\lambda$ . **d**. Temperature profile reached after a 7.5 V and 13 µs pulse in a graphene microheater with a $w = 10$ $\mu m$ bridge, calculated using our 3D COMSOL Multiphysics model. A 30-nm-thick, 3 µm-long, and 800- nm-wide GSST cell was considered, enough to introduce $\pi$ phase-shift upon switching from the amorphous to the crystalline state. The dotted lines delimit the single-layer graphene sheet. **e.** Temperature evolution as a function of time for the device shown in **d** for two 7.5 V pulses with 5 µs and 13 µs durations. Similar to Figure 3, the cooling process was fitted best with a double exponential function featuring two thermal constants.

### 3.2 Nonvolatile reconfigurable transmissive metasurfaces

We propose a transparent reconfigurable metasurface based on GSST and a single-layer graphene microheater. We take advantage of the broadband transparency of graphene and its two-dimensionality to build a microheater that enables reconfiguration with no perturbation to the optical resonance modes. The metasurface consists of a 2D array of GSST cylindrical meta-atoms which introduce a phase-delay that depends on the refractive index dictated by the material's structural state. We use the experimental values of the refractive index obtained in Ref. [39] and assume that any intermediate value between the crystalline and the amorphous states can be reached.[45] To optimize the meta-atom geometry, we utilized a deep neural network design framework demonstrated in Ref [79]. The neural network was set to ensure close-to-2π phase coverage and to maintain high transmittance for GSST meta-atoms constrained to a maximum thickness of 400 nm. The optimized design parameters are: cylinder radius – 445 nm, cylinder height – 390 nm, square lattice period – 1.35 µm, and wavelength – 2.21 µm. The effect of graphene on the meta-atom geometry was simulated in COMSOL Waves & Optics Module as a conducting surface with $\sigma = (5.85-1.01i) \times 10^{-5}$ S for Fermi level $E_f = 0.2$ eV, and $\sigma = (5.97-3.10i) \times 10^{-6}$ S for $E_f = 0.34$ eV.[80,81] Figure 5a shows the distinct electromagnetic fields in fully amorphous and fully crystalline GSST meta-atoms. To simulate the intermediate states, we estimated the refractive indices assuming different densities of crystalline nuclei embedded in amorphous GSST and employing the Lorentz-Lorentz equation for effective medium to retrieve the corresponding intermediate refractive indices.[82] We demonstrate in Figure 5b a 294° phase range modulation with nearly uniform amplitude by finely controlling the crystalline fraction of GSST. We compare this result to the calculation of the same meta-atom without graphene to find that the amplitude and phase are unaltered by the graphene layer (for $E_f = 0.34$ eV or higher). This result demonstrates the non-perturbative nature of the graphene heater. Furthermore, we plot the phase modulation for two different graphene Fermi levels; even under different charge doping concentrations, the phase modulation is near identical and only the amplitude decreases slightly due to the higher losses with lower chemical potentials ($E_f = 0.2$ eV in this case).

We now study the thermal conditions to switch the entirety of the meta-atoms. Since the GSST thickness is 390 nm, it is crucial to guarantee uniform switching throughout the meta-atom thickness. We found that the long pulse sequence used so far in this work (See Figure 2) favors uniform switching of the entire meta-atom. If the power is enough to reach the transition temperature, approximately 5 µs are required for both the upper and bottom surfaces of the meta-atom to reach the same temperature during the heating process. Figure 5c shows the temporal evolution of the average temperature of both surfaces for the 13 µs amorphization pulse and the 20 ms crystallization pulse in a 25 × 25 µm² square microheater. In general, the top surface heats up and cools down slower than the bottom one; this occurs due to the weaker heat transfer with the surrounding air than with the substrate. During the cooling process, the temperature gradient reaches a maximum of 50 K difference between both surfaces, and in nearly 500 ns longer time to quench. However, we anticipate that this effect will not affect the re-amorphization process of GSST, which exhibits slower crystallization kinetics and larger critical thickness compared to classical GST and, therefore, features significantly larger reversible switching volume.[45] In the case of the crystallization pulse, for which we use a 1.5 ms trailing edge, the entire meta-atom is kept at the almost same temperature during the cooling process, with just 3 K difference between the bottom and top surfaces. Hence, the trailing edge effectively eliminates temperature gradients, thus enabling uniform distribution of nuclei in the entire volume. This effect is critical to controlling intermediate levels, which would only depend on pulse power and duration with guaranteed heating uniformity.

In Figs. 5d and 5e we plot the voltage and power required to reach a temperature of 890 K with a square single-layer graphene microheater, considering again $h_{SiO/Gr} = 1.5 \times 10^5$ W m$^{-2}$ K$^{-1}$. If $P_{Am}$ is the power dissipated only by the graphene microheater, which is required to amorphize the entire 390-nm-thick GSST, then the total voltage to reach $P_{Am}$ is given by $V = R_T\sqrt{P_{Am}/R_{sh}}$, where $R_T = 2R_c + R_{sh}$ is the total resistance. In Figure 5e, we show the results for $P_{Am}$ calculated using our 3D model and fitted with a quadratic function for different heater sizes and employing the same 13 μs and 20 ms pulses for amorphization and crystallization, respectively. Because $P_{Am}$ is constant for a given graphene microheater area, we can estimate the total voltage as a function of both sheet and contact resistances, as shown in Figure 5d. In particular, we find that the contact and sheet resistance measured in our experiments (see Figure 1) lead to a power $P_{Am} = 1.34$ mW and a voltage $V_1 = 2.5$ V to actuate switching across a $1 \times 1$ μm$^2$ microheater, which fits a single meta-atom. Similarly, $P_{Am} = 1.24$ W and a voltage $V_{100} = 77.8$ V are required to actuate switching across a $100 \times 100$ μm$^2$ microheater, which fits a $75 \times 75$ meta-atom array. The microheater can thus be configured for either collective switching [83] of metasurfaces or individual meta-atoms tuning.[84]

Several improvements can be made to enhance the heater performance further. To minimize the contact resistance between electrodes and graphene, a 1D edge coupling approach [85] or patterning the electrodes [86] could reduce the total voltage required to dissipate $P_{Am}$ on the heater. Morever, to avoid the current saturation and reach high temperatures, multi-layer graphene could be used,[87] and hexagonal boron nitride encapsulated graphene devices can be employed as they sustain high lattice temperature up to ∼1600 K.[64] Finally, to reduce the sheet resistance and also avoid the current saturation, a gate terminal could be added to control the charge doping of graphene.[58]

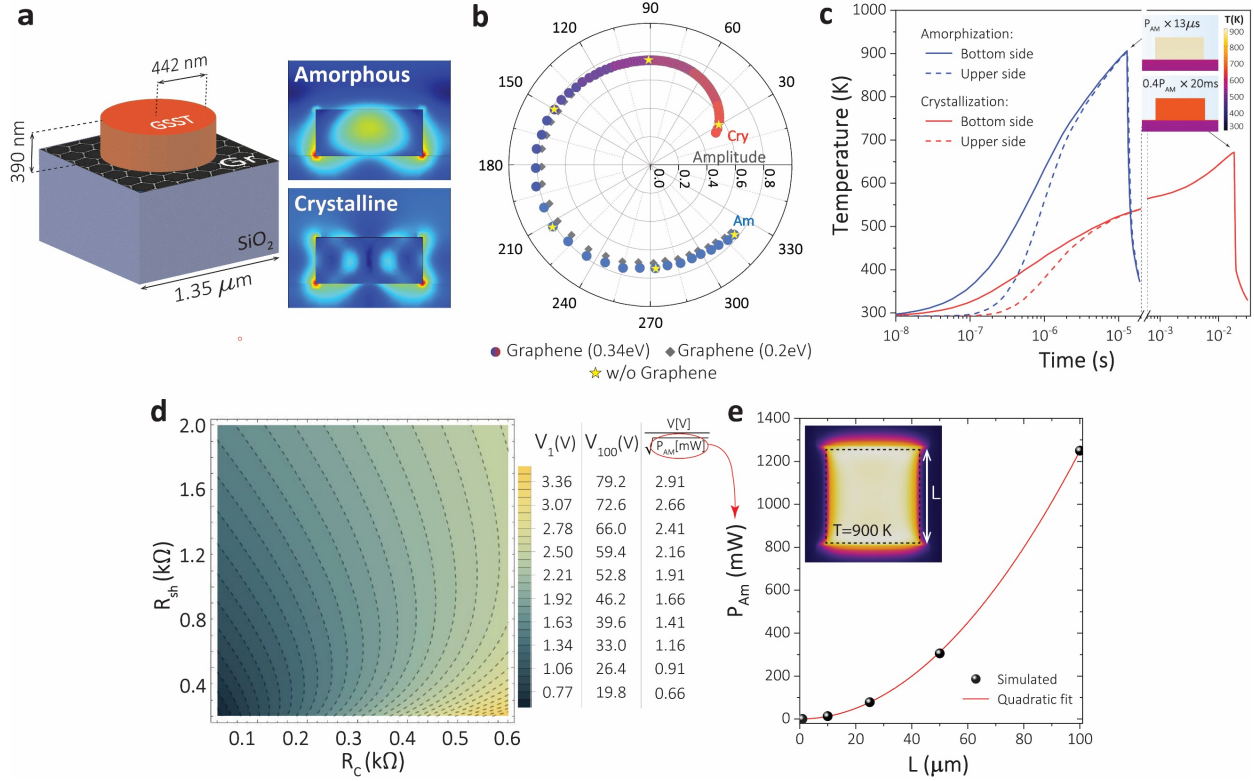

**Figure 5. GSST-based metasurface with non-perturbative graphene microheater. a** Illustration of GSST meta-atom on a graphene microheater. The meta-atom geometry was optimized for operational wavelength λ = 2.21 μm by exploiting a deep neural network design framework.[79] The two diagrams on the right depict the E-field magnitude distribution inside the meta-atoms in fully amorphous and crystalline states; graphene chemical potential $E_f$ = 0.34 eV. **b** Phase and amplitude modulation of the meta-atom plotted in **a** with and without graphene. Grey diamonds and colored circles correspond to the meta-atom with 0.2 eV and 0.34 eV graphene, respectively. Yellow stars indicate a group of 8 selected meta-atoms with equidistant phase-delay steps of 45° when no graphene is used. **c** Temperature evolution during amorphization and crystallization, with pulse powers $P_{Am}$ and $0.4\ P_{Am}$ and durations of 13 μs and 20 ms, respectively. The inset shows uniform temperature distribution across the meta-atom. The bottom and upper sides reach the same temperature within approximately 5 μs. **d** Voltage required to switch a square microheater of side L as a function of the contact, $R_c$, and sheet resistances $R_{Sh}$. This voltage is given by the function $V = (2R_c + R_{sh})\sqrt{P_{Am}/R_{sh}}$, and depends on the power required to amorphize, $P_{Am}$, $V_1$ and $V_{100}$ are the voltages required to switch with a square microheater of side L = 1 μm (for a single meta-atom) and L = 100 μm (for an array of 75 × 75 meta-atoms), respectively. **e** $P_{Am}$ calculated as a function of L, the length of the side on the square microheater. The quadratic fitting gives $P_{Am}$=0.124 + 1.25·$L^2$.

## 4. Conclusion

In this paper, we demonstrated, for the first time, reversible electro-thermal switching of PCM using a single-layer graphene microheater, with up to 4 distinct levels and low measured power of 8.6 ± 1.1 mW. Our result positions single-layer graphene as the lowest power consumption approach for switching PCMs of comparable size. As a side effect, we observed a current saturation phenomenon resulting from SPoPh scattering that limits the power that can be fed to the system.

Additionally, we built a full 3D multi-physics model to precisely reproduce our experimental results. We found that approximating the effect of the electronic and phononic phenomena to the thermal conductivity of the graphene/SiO$_2$ boundary, and assuming a constant contact and sheet resistances gives accurate results since we observed good agreement between simulation and experiments. A more realistic simulation, at the cost of computational time, should take into account the real-time variation of the properties as a function of temperature and voltage. We used our experimental and computation results to reveal the technical limitations when using single-layer graphene on SiO$_2$ substrate, and in turn inform the design of photonic devices based on this platform. In particular, we proposed a design to achieve low-loss phase modulation in PCM-integrated SiN waveguides with an ultra-compact form factor. Moreover, we demonstrated reconfigurable GSST meta-atoms capable of tuning the optical phase across a 294° range, which can serve as a building block to construct active metasurfaces capable of arbitrary wavefront modulation. The broadband transparency, substrate-agnostic integration capability, and minimal perturbation on optical modes characteristic of the graphene heater, together with the nonvolatile nature of PCM qualifies this platform as a promising solution to next-generation reconfigurable free-space and integrated optical devices.

## 5. Experimental section

*Device Fabrication:* We transferred chemical vapor deposition (CVD) grown single-layer graphene onto 3-μm-thick SiO$_2$/Si wafers following the standard wet transfer technique.[88] We patterned the graphene using electron beam lithography with Ma-N 2403 negative photoresist (Microresist technology), followed by O$_2$ plasma etching. Subsequently, we added 100 nm Ti/Au contacts using a second electron beam patterning on polymethyl methacrylate (PMMA) photoresist, electron beam metal evaporation, and a lift-off process. Next, we deposited 10 nm protective Al$_2$O$_3$ layers using atomic layer deposition. Lastly, we used a third electron beam lithography step to pattern PMMA windows, followed by thermal evaporation of GSST and a final lift-off process. Another 10 nm of Al$_2$O$_3$ were deposited to protect GSST from oxidation. We fabricated graphene microheaters consisting of two $27.5 \times 100$ μm$^2$ graphene pads connected by bridges with widths of 3, 5, and 10μm and a fixed length of 5 μm. The metal Ti/Au pads were in contact with a total of $17 \times 100$ μm$^2$ graphene on each side to minimize the contact resistance.

*Experimental setup:* We wire-bonded the chip onto a custom PCB for easy electrical testing, and mounted the device onto a customized 3D printed stage compatible with a Renishaw Invia Reflex Micro Raman System.[65] We carried out *in-situ* Raman probing using a 785 nm laser, a 1200 lines per millimeter grating, and a 100 × long working-distance objective. We used a line focusing approach to obtain the Raman signal from the entire GSST cell, instead of focusing on a spot. The electrical pulses were sent using an analog 1 GHz, 20 V pulse generator with a minimum of 0.5 ns raising and trailing edges.

*Computational modelling:* To simulate heat generation by Joule heating and its dissipation, we used COMSOL Multiphysics. We used the module *Electric Currents, Shell* to approximate the simulation of graphene as a boundary with a thickness $d = 0.34$ nm. We coupled this module with the *Heat Transfer in Solids* module, where surface-to-surface radiation and thermal boundary resistance were considered. In particular, we used a thermal boundary conductivity of $10^8$ W m$^{-2}$ K$^{-1}$ for all interfaces, with no considerable effects, except for the boundary between graphene and SiO$_2$, $h_{SiO/Gr}$ which was extensively studied in Figure 2. To carry out the simulations, we used the following parameters for the materials involved:

|  | Si | SiO$_2$ | Si$_3$N$_4$ | am-GSST | cry-GSST | Graphene | Al$_2$O$_3$ | Au |
|---|---|---|---|---|---|---|---|---|
| **Density (kg/m$^3$)** | 2329 | 2203 | 3100 | 5267[a] | 5267[a] | 2250 | 3900 | 19300 |
| **Specific heat (J/(kg K))** | 700 | 740 | 700 | 275 | 351 | 420 | 900 | 129 |
| **Thermal conductivity (Wm$^{-1}$K$^{-1}$)** | 150 | 1.38 | 20 | 0.2 | 0.4 | 160 [89][b] | 30 | 317 |
| **Relative Permittivity** | - | - | - | - | - | 4.708 | - | 6.9 |
| **Electrical conductivity (S/m)** | - | - | - | - | - | $1/(d \cdot R_{sh})$ | - | 45.6×10$^6$ |

[a] Estimation using molar density [b] considering both SiO$_2$ substrate and devices with ~5 $\mu m$ features, which decrease graphene thermal conductivity.

Although our simplified simulation does not consider temperature dependence of the thermal and electrical conductivity of graphene and phase transition of GSST, the model reproduced the results obtained in the experiment with remarkable accuracy.

To calculate the phase and amplitude modulation by the metasurface in Figure 5, we used the COMSOL Waves & Optics Module. A single meta-atom with Bloch boundary conditions was simulated using the refractive index of GSST from Ref. [39] and $\sigma = (5.85-1.01i) \times 10^{-5}$ S for graphene with $E_f = 0.2$ eV, and $\sigma = (5.97-3.10i) \times 10^{-6}$ S for $E_f = 0.34$ eV, at $\lambda = 2.21$ $\mu$m.[80,81] The mode calculations shown in Figure 4 were carried out in Lumerical, considering the same GSST refractive index and the built-in refractive indices for Si, Al$_2$O$_3$, SiO$_2$, Si$_3$N$_4$ and gold. The surface conductivity of graphene was calculated with Lumerical's built-in function with chemical potentials of 0.2 eV and 0.34 eV. All simulations accounted for the 10 nm alumina, which is transparent in the 1-17 μm wavelength window, and served both as a protective layer and as a spacer to mitigate the large temperature gradient between the graphene/$SiO_2$ interface and GSST.

**Acknowledgements**
This material is based upon work supported by the DARPA Young Faculty Award Program under Grant Number D18AP00070 and the Assistant Secretary of Defense for Research and Engineering under Air Force Contract No. FA8702-15-D-0001. M.S., T.G., K.R., M.K., and J.H. also acknowledge funding support provided by the DARPA Extreme Optics and Imaging (EXTREME) Program under Agreement No. HR00111720029. The authors acknowledge fabrication facility support by the MIT Microsystems Technology Laboratories and the Harvard University Center for Nanoscale Systems, the latter of which is supported by the National Science Foundation under award No. 0335765.